\documentclass[12pt]{article}

\usepackage[dvips]{color}
\usepackage{epsfig}
\def\lsim{\mathrel{\rlap{
\lower4pt\hbox{\hskip-3pt$\sim$}}
    \raise1pt\hbox{$<$}}}     %less than approx. symbol
\def\gsim{\mathrel{\rlap{
\lower4pt\hbox{\hskip-3pt$\sim$}}
    \raise1pt\hbox{$>$}}}     %greater than or approx. symbol

\newcommand{\bc}{\begin{center}}
\newcommand{\ec}{\end{center}}
\newcommand{\be}{\begin{equation}}
\newcommand{\ee}{\end{equation}}
\newcommand{\bea}{\begin{eqnarray}}
\newcommand{\eea}{\end{eqnarray} }
\newcommand{\ba}{\begin{array}}
\newcommand{\ea}{\end{array}}

\begin{document}
%\section{draft}
\begin{center}
{\bfseries  OBSERVABLE CONSEQUENCES \\ OF CROSSOVER-TYPE
DECONFINEMENT PHASE TRANSITION}

\vskip 5mm

 V.D.~Toneev$^{\dag}$

\vskip 5mm

{\small  {\it Bogoliubov Laboratory of Theoretical Physics Joint
Institute for Nuclear Research, \\ Dubna, Russia }
\\
$\dag$ {\it E-mail: toneev@thsun1.jinr.ru }}
\end{center}

\vskip 5mm

\begin{center}
\begin{minipage}{150mm}
\centerline{\bf Abstract} Equation of State (EoS) for hot and
dense nuclear matter with a quark-hadron phase transition is
constructed within a statistical mixed-phase model assuming
coexistence of unbound quarks in nuclear surrounding. This model
predicts the de\-con\-finement phase transition of the crossover
type.
 The so-called "softest point" effect of EoS is analyzed and
confronted to that for other  equations of state which exhibit the
first order phase transition (two-phase bag model) or no
transition at all (hadron resonance gas). The collective motion of
nucleons from high-energy heavy-ion collisions is considered
within a relativistic two-fluid hydrodynamics for different EoS.
It is demonstrated that the  beam energy dependence
 of the  directed flow is
a smooth function in the whole range from SIS till SPS energies
and allows to disentangle  different EoS,  being in good agreement
with experimental data for the statistical mixed-phase model. In
contrast, excitation functions for relative stran\-ge\-ness
abundance turn out to be insensitive to the order of phase
transition.
\\[3mm]
\end{minipage}
\end{center}

\vskip 10mm
\section{Introduction}
The predicted phase transition from confined hadrons to a
deconfined phase of
 their constituents ({\it i.e.} the asymptotically-free quarks and gluons
or the so-called Quark-Gluon Plasma, QGP) is a challenge to the
theory of strong interaction. Over the past two
 decades a lot of efforts has been spent to both the theoretical study of
 deconfinement phase transition and the
 search for its possible  manifestation
 in relativistic heavy ion collisions, properties of neutron stars and
 Universe evolution. A unique opportunity provided by relativistic heavy-ion
 collisions allowed to reach a state with  temperature and  energy density
exceeding the  critical values, $T_c\sim 170 \ MeV$ and
$\varepsilon_c \sim 1 \ GeV/fm^3$, specific for the deconfinement
phase transition. A rather long list of various signals for the
QGP formation in hot and dense nuclear matter is available now and
it has been probed in experiments with heavy ions. Unfortunately,
there is no crucial signal for unambiguous identification of the
deconfinement phase and, for a particular reaction at the given
bombarding energy,
 practically   every proposed signal can be simulated to some extent by
 hadronic interactions.

In this paper we turn to the study of excitation functions for
observables to be sensitive to the expected QCD deconfinement
phase transition. Its manifestation has been considered already
some time ago by~\cite{SZ89,V83}. Since a phase transition slows
down the time evolution of the system due to {\em softening} of
the EoS, and one can expect  a remarkable loss of correlations
around some critical incident energy resulting in  definite
observable effects.

\section{Equation of state in mixed phase model}

Following the common strategy of the two-phase (2P) bag
model~\cite{Cleym}, one can determine the deconfinement phase
transition by means of the Gibbs conditions matching the EoS of
 a relativistic gas of hadrons and resonances, whose interactions  are
simulated by the Van der Waals excluded volume correction, to that
of an ideal gas of quarks and gluons, where the change in vacuum
energy in a QGP state is parameterized by the bag constant $B$.
Thermodynamics of the hadron gas is described in the grand
canonical ensemble.  All hadrons with the mass $m_j < 1.6 \ GeV$ i
are taken into consideration. One should emphasize that the phase
transition in the 2P model is right along of the first order by
constructing.

To reproduce the variety of phase transitions predicted by QCD
lattice calculations we represent a phenomenological Mixed Phase
(MP) model~\cite{NST98,TNS98}. The underlying assumption of the MP
model is that unbound quarks and gluons {\it may coexist} with
hadrons forming a {\it homogeneous} quark/gluon--hadron phase.
Since the mean distance between hadrons and quarks/gluons in this
mixed phase may be of the same order as that between hadrons, the
interaction between all these constituents (unbound quarks/gluons
and hadrons) plays an important role and defines the order of the
phase transition.

Within the MP model \cite{NST98,TNS98} the effective Hamiltonian
is expressed in the quasiparticle approximation with
density-dependent mean-field interactions.
 Under quite general requirements of  confinement
for color charges, the mean-field potential of  quarks and gluons
is approximated by
\begin{equation}
U_q(\rho)=U_g(\rho)={A\over\rho^{\gamma}}~;
 \ \ \ \gamma >0
\label{eq6}
   \end{equation}
with  {\it the total density of quarks and gluons}
$$
\rho=\rho_q + \rho_g +\sum\limits_{j}\;\nu_j\rho_{j}~,
$$
where $\rho_q$ and  $\rho_g$ are the densities of unbound quarks
and gluons outside of hadrons, while $\rho_{j}$ is the density of
hadron type $j$ and $\nu_j$ is the number of valence quarks
inside. The presence of the total density $\rho$ in (\ref{eq6})
implies interactions between all components of the mixed phase.
The  approximation (\ref{eq6})  mirrors two important limits of
the QCD interaction. For
 $\rho \to 0$, the interaction potential approaches infinity,
{\em i.e.}  an infinite energy is necessary to create an isolated
quark or gluon, which simulates the confinement of color objects.
In the other extreme case of large energy density corresponding to
$\rho \to \infty$, we have  $U_q=U_g=0$ which is consistent with
asymptotic  freedom.

The use of the density-dependent  potential (\ref{eq6}) for quarks
 and the hadronic potential, described by a
modified non-linear mean-field model~\cite{Zim},  requires certain
constraints to be fulfilled, which are related to thermodynamic
consistency~\cite{NST98,TNS98}. For the chosen form of the
Hamiltonian these conditions require that  $U_g(\rho)$ and
$U_q(\rho)$ do not depend on temperature. From these conditions
one also obtains a form for the quark--hadron
potential~\cite{NST98}.

A detailed study of the pure gluonic $SU(3)$ case with a
first-order phase transition allows one to fix the values of the
parameters as $\gamma =0.92$ and $\displaystyle A^{1/(3\gamma+1)}
= 250$ MeV. These values are then used for the
 $SU(3)$ system including quarks. As is shown in Fig.1
for the case of quarks of two light flavors at zero baryon density
($n_B=0$), the MP model is consistent with lattice QCD data
providing a continuous phase transition if the crossover   type
with a deconfinement temperature $T_{dec}=153$ MeV.
 For a two-phase approach based on the bag model  a first-order
deconfinement phase transition occurs with a sharp jump in energy
density $\varepsilon$ at $T_{dec}$ close to the value obtained
from lattice QCD.

%\vspace*{-2mm}
\begin{figure}%[htb]
\begin{center}
\leavevmode \epsfxsize=10.cm \epsfbox{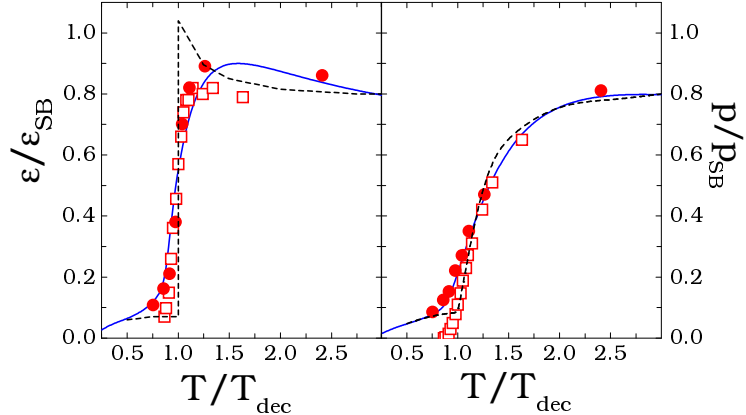}
\end{center}
\vspace{-8mm}

\caption[C1]{The reduced energy density  and pressure  (the
$\varepsilon_{SB}$ and $p_{SB}$ are corresponding
Ste\-phan-Boltz\-mann quantities)
 of the $SU(3)$ system with two light flavors for $n_B=0$ calculated
within the MP (solid lines) and bag (dashed lines) models. Circles
and  squares are  lattice QCD data obtained within the
Wilson~\cite{RS86} and Kogut--Susskind~\cite{berna} schemes.  }
  \label{fig1}
\end{figure}

%\vspace{-5mm}

Though at a glimpse the temperature dependencies of the energy
density $\varepsilon$ and pressure $p$ for the different
approaches presented in Fig.1 look quite similar, there is a large
difference revealed when $p/\varepsilon$ is plotted versus
$\varepsilon$ (cf. Fig.2, left panel).  The lattice QCD data
differ at low $\varepsilon$, which is due to difficulties within
 the Kogut--Susskind scheme~\cite{berna} in treating the
hadronic sector. A particular feature in the MP model is that, for
$n_B=0$, the {\em softest point} of the EoS, defined  as a minimum
of the function  $p(\varepsilon)/\varepsilon$ \cite{HS95}, is not
\begin{figure}%[h]
\begin{center}
\leavevmode \epsfxsize=10.cm \epsfbox{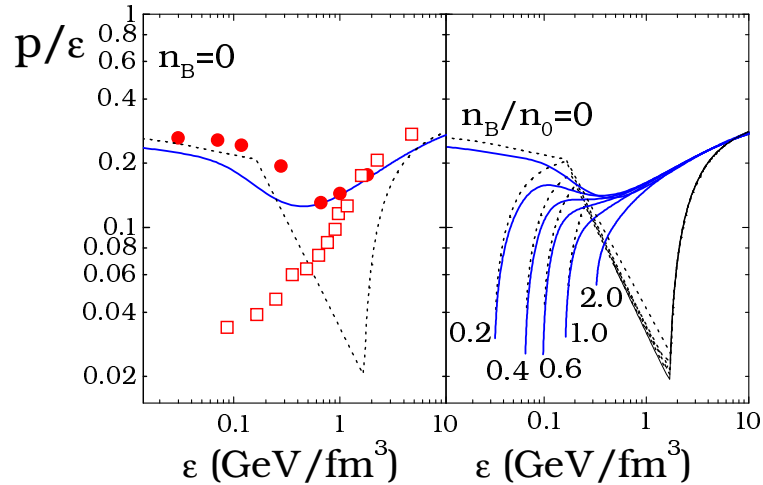}
\end{center}
\vspace{-8mm}
 \caption[C2]{The ($\varepsilon
,p/\varepsilon$)-representation of the EoS for the two-flavor
$SU(3)$ system at various baryon densities $n_B$. Notation of data
points and lines is the same as in Fig.1. }
\label{fig2}
\end{figure}
very pronounced and located at comparatively  low values of the
energy density: $\varepsilon_{SP} \approx 0.45$ GeV/fm$^3$, which
roughly agrees with the lattice QCD value~\cite{RS86}.  This value
of $\varepsilon $ is close to the energy density inside a nucleon,
and hence, reaching this value indicates that we are dealing with
a single {\em big hadron} consisting of deconfined matter. In
contradistinction, the bag-model EoS exhibits a very pronounced
softest point at large energy density $\varepsilon_{SP} \approx
1.5$ GeV/fm$^3$~\cite{HS95,R96}.

The MP model can be  extended to baryon-rich  systems in a
parameter-free way \cite{NST98,TNS98}. As demonstrated in Fig.2
(right panel), the softest point for baryonic matter is gradually
washed out with increasing  baryon density and vanishes for  $n_B
\gsim 0.3 \ n_0$ ($n_0$ is normal nuclear matter density). This
behavior differs drastically from  that of the two-phase bag-model
EoS, where $\varepsilon_{SP}$ is only weakly dependent on
$n_B$~\cite{HS95,R96}. It is of interest to note that
 the interacting hadron gas model has  no softest point at all and,
in this respect, its thermodynamic behavior is close to that of
the  MP model at high energy densities~\cite{TNS98}.

 These differences between the various models of EoS should manifest
themselves in  dynamics discussed below.

\section{Directed flow of baryons}
The EoS described above is applied to a two-fluid (2F)
hydrodynamic model~\cite{MRS91}, which takes into account finite
stopping power of  colliding heavy ions. In this dynamical model,
the total baryonic current and energy-momentum tensor are written
as
   \begin{eqnarray}
J^{\mu} &=& J^{\mu}_p + J^{\mu}_t~~,
   \label{eq7.6}
\\
T^{\mu\nu} &=& T^{\mu\nu}_p + T^{\mu\nu}_t~~,
   \label{eq7}
   \end{eqnarray}
where  the baryonic current
$J^{\mu}_{\alpha}=n_{\alpha}u_{\alpha}^{\mu}$ and energy-momentum
tensor $T^{\mu\nu}_{\alpha}$  of the fluid $\alpha$ are initially
associated with either target ($\alpha=t$) or projectile
($\alpha=p$) nucleons. Later on
 these fluids contain all hadronic and quark--gluon species,
depending on the model used for describing the fluids. The twelve
independent quantities (the baryon densities $n_{\alpha}$,
4-velocities  $u_{\alpha}^{\mu}\;$ normalized as
  $u_{\alpha\mu}u_{\alpha}^{\mu}=1$,
as well as temperatures $T$ and pressures $p$ of the fluids) are
obtained by solving the following set of equations of two-fluid
hydrodynamics~\cite{MRS91}
   \begin{eqnarray}
   \label{eq8}
   \partial_{\mu} J_{\alpha}^{\mu} &=& 0~~, \\
   \partial_{\mu} T^{\mu\nu}_{\alpha} &=& F_{\alpha}^\nu~~,
   \label{eq8a}
   \end{eqnarray}
where the coupling term
 \begin{eqnarray}
 F_{\alpha}^\nu=n^s_p n^s_t
\left < V_{rel} \int d\sigma_{NN\to NX}(s)\; (p - p_{\alpha})^\nu
\right >
 \label{eq9}
\end{eqnarray}
characterizes friction between the  counter-streaming fluids.
 The cross sections $d\sigma_{NN\to
NX}$ take into account all elastic and inelastic interactions
between the  constituents of different fluids at the invariant
collision energy $s^{1/2}$ with the local relative velocity
$V_{rel} =[s(s-4m_N^7)]^{1/2}/2m_N^2~.$ The average in (\ref{eq9})
is taken over  all particles in the two  fluids which are assumed
to be in local equilibrium intrinsically~\cite{MRS91}. The set of
Eqs.~(\ref{eq8}) and (\ref{eq8a}) is closed by  EoS, which is
naturally the same for both colliding fluids.

Following the original paper~\cite{MRS91}, it is assumed
 that a fluid element decouples from the hydrodynamic regime,
when its baryon density $n_B$ and densities in the eight
surrounding cells become smaller than a fixed value  $n_f$. A
value $n_f = 0.8 n_0$ is used for this local freeze-out density
which corresponds to the actual density of the freeze-out
 fluid element of about $0.6-0.7\ n_0$.

The directed  flow characterizes the deflection of emitted hadrons
away from the beam axis within the reaction $x-z$ plane. In
particular, one defines the differential directed flow by the mean
in-plane component $\left< p_x(y)\right>$ of the transverse
momentum at a given rapidity $y$.  This deflection is believed to
be quite sensitive to the {\em elasticity} or {\em softness} of
the EoS and can be quantified in two ways: In terms of the
derivative (a slope parameter) at mid-rapidity
\begin{equation}
\label{eq12} F_y = \left. \frac{d \; \left<p_x(y)\right>}{dy}
\right|_{y=y_{cm}}~~,
\end{equation}
which is quite suitable for analyzing the flow excitation
function, and by another integral quan\-tity to be less sensitive
to possible rapidity fluctuations of the in-plane momentum:
\begin{equation}
\left< P_x\right> = \frac{\displaystyle\int dp_xdp_ydy \ p_x \
\left(  E{\displaystyle
\frac{d^3N}{dp^3}}\right)}{\displaystyle\int  dp_xdp_ydy \ \left(
    E{\displaystyle
 \frac{d^3N}{dp^3}}\right) }~~,
\label{eq13}
\end{equation}
where the integration in the c.m.system runs over the rapidity
region $\displaystyle [0,y_{cm}]$. Excitation functions in the
SIS-AGS-SPS energy range are plotted in Fig.3 for both
characteristics~\cite{INNST02}.

\begin{figure}[htb]
\label{fig3}
\begin{center}
%\leavevmode \epsfxsize=11.cm \epsfbox{fig3_fl.eps}
  \epsfig{file=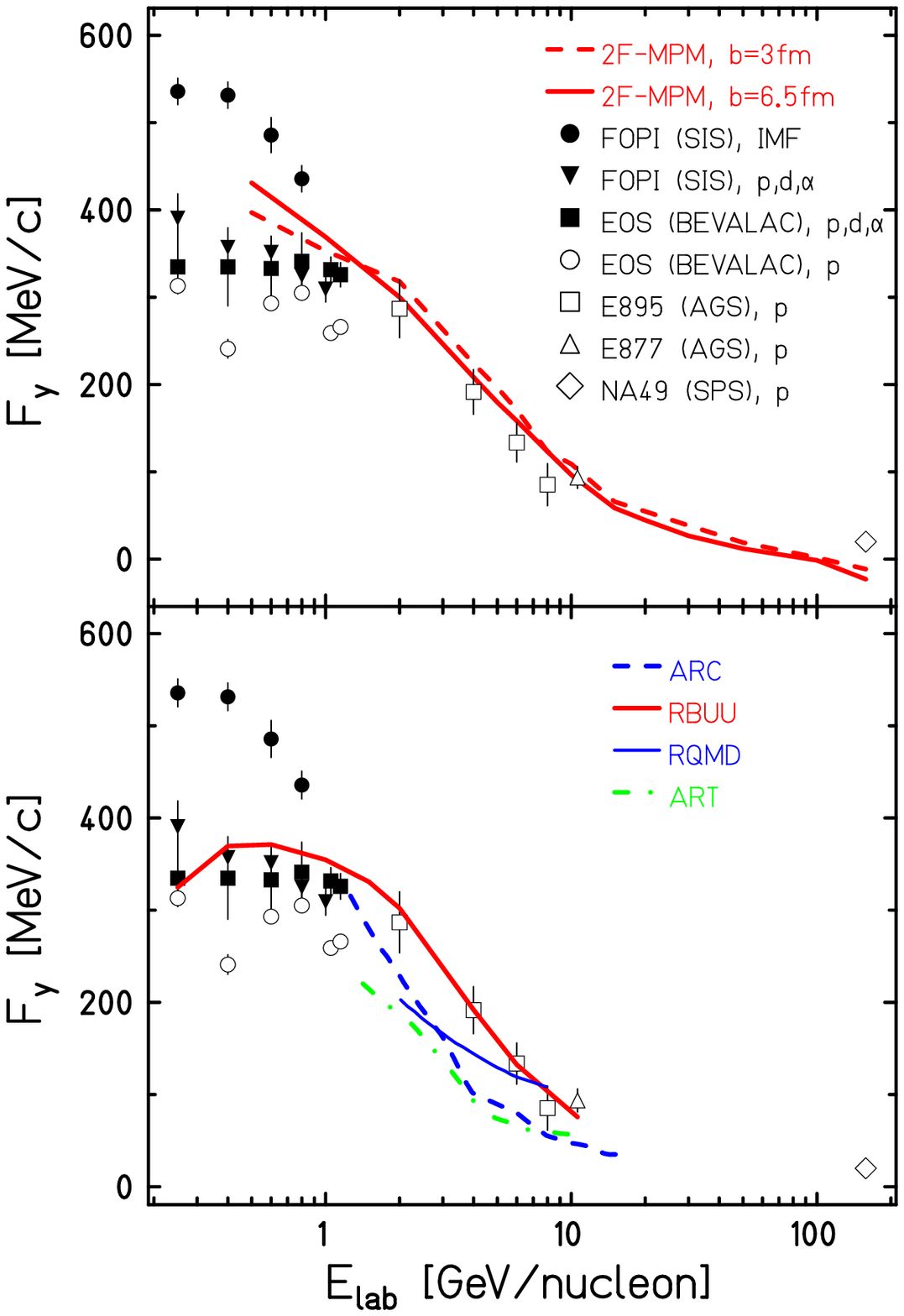, width=0.4\textwidth%, height = 75\unitlength
  } \hspace*{10mm}
  \epsfig{file=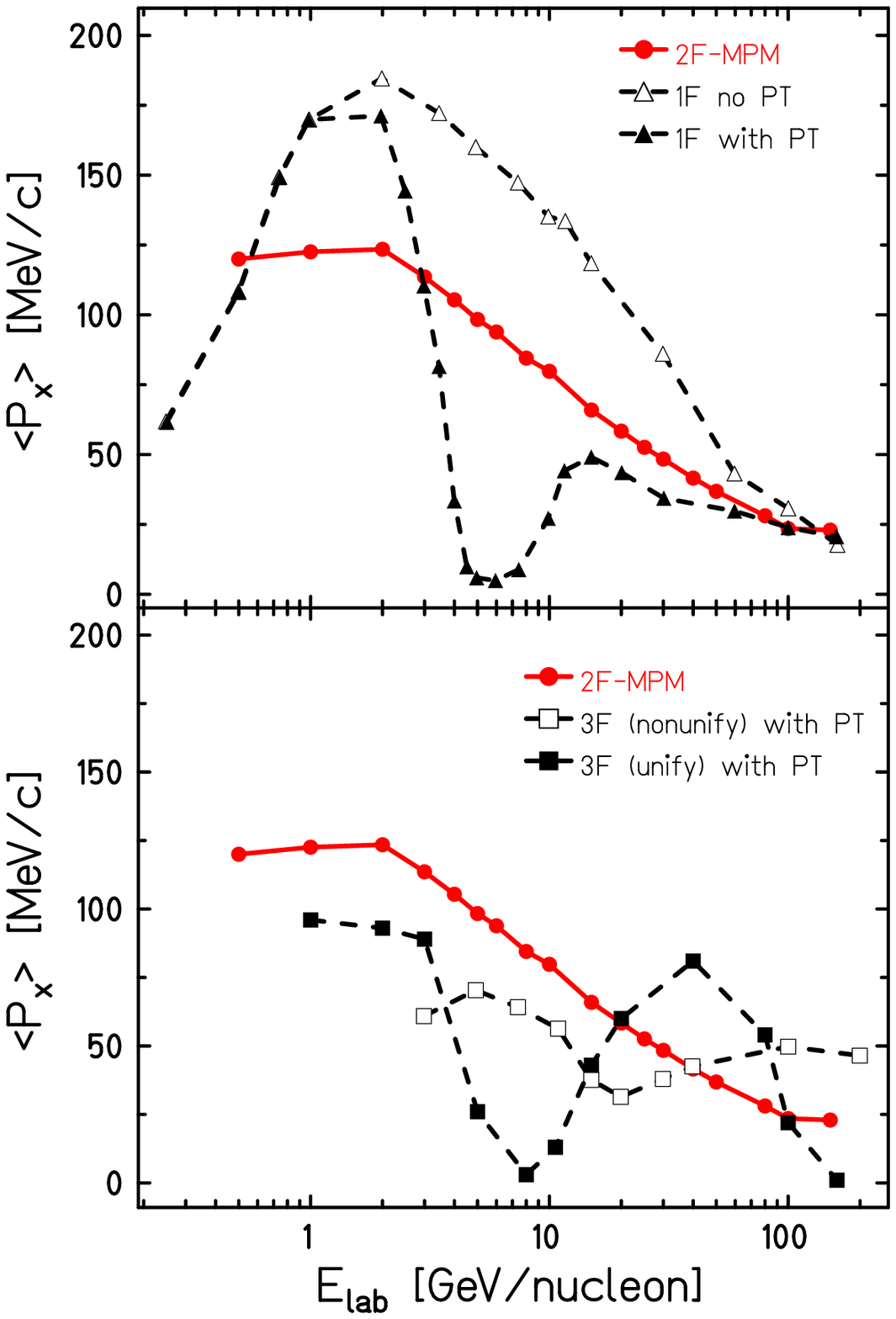, width=0.4\textwidth%, height = 75\unitlength
  }
\end{center}
\vspace{-8mm}

 \caption[C3]{Excitation functions of the slope
parameter $F_y$ (left panel) and the average directed flow (right
panel) for baryons from Au + Au collisions within hydrodynamics
and different transport simulations. Collected experimental points
are taken from~\cite{INNST02}. The results of transport
calculations for three different codes (left lower panel) are
given by the thin solid (RQMD), dashed (ARC) and dot-dashed (ART)
lines (cited according to~\cite{Aj98}). The solid line (RBUU) is
taken from~\cite{cas99}.
 2F hydrodynamics with the MP EoS at the impact
parameter 3 fm is compared with the corresponding results of
1F-~\cite{R96} (right upper panel) and 3F- (right lower
panel)~\cite{3DF} hydrodynamics with the bag-model EoS. 1F
calculations both with and without the phase transition (PT) are
displayed.  }
\end{figure}

Our first 2F hydrodynamic calculations of $F_y (E_{lab})$ are in a
good agreement with experiment in the whole energy range
considered. In the left lower panel of Fig.3  our results  are
compared with transport calculations. The ARC and ART are cascade
models, while the RQMD takes also into account mean-field effects.
Though all these models agree with experimental data at $E_{lab}
\approx 10$ A$\cdot$GeV (considered as a reference point), values
of $F_y$ at lower energies are clearly underestimated, as is
evident from comparison with results of the E895
Collaboration~\cite{E895} (see empty squares in Fig.3). Recently,
a good description of experimental points (including the E895
data) was reported within a relativistic BUU  (RBUU)
model~\cite{cas99}. The good agreement with experiment was
achieved here by a special fine tuning of the mean fields involved
in the particle propagation.

The calculated excitation functions of $\left< P_x\right>$  for
baryons within different hydrodynamic models are shown in the
right panel of Fig.3. Conventional 1F hydrodynamics
 for pure hadronic matter~\cite{R96} results
 in a  very large directed flow due to the inherent
instantaneous stopping of the colliding matter. This instantaneous
stopping is unrealistic at high beam energies. If the
 deconfinement phase transition, based on the bag-model EoS~\cite{R96},
is included into this model, the excitation function of $\left<
P_x\right>$ exhibits a deep minimum near $E_{lab}\approx 6$
A$\cdot$GeV, which manifests the softest-point effect of the
bag-model EoS as shown in the right panel of Fig.2.

The result of 2F hydrodynamics with the MP EoS noticeably differs
from the 1F calculations. After a maximum around 1 A$\cdot$GeV,
the average directed flow  decreased slowly and smoothly. This
difference is caused by two reasons. First, as follows from Fig.2,
the softest point of the MP EoS is washed out for $n_B \gsim 0.4$.
The second reason is dynamical: the finite stopping power and
direct pion emission change the evolution pattern. The latter
point is confirmed by comparison to three-fluid calculations with
the bag EoS~\cite{3DF} plotted in the right lower panel of Fig.3.
 The third pionic fluid in this model
is assumed to interact only with itself neglecting the interaction
with baryonic fluids. Therefore, with regard to the baryonic
component, this three-fluid hydrodynamics~\cite{3DF,3DF97} is
completely equivalent to our two-fluid model and the main
difference is due to the different EoS. As seen in Fig.3, the
minimum of the directed flow excitation function, predicted by the
one-fluid hydrodynamics with the bag-model   EoS, survives in the
three-fluid (nonunified) regime but its value decreases and its
position shifts to higher energies. If one  applies the
unification procedure  of ~\cite{3DF}, which favors fusion of two
fluids into a single one, and thus making stopping larger,
three-fluid hydrodynamics practically reproduces  the one-fluid
result and predicts in addition a bump at $E_{lab}\approx 40$
A$\cdot$GeV.

\section{Strangeness production}
Enhanced strangeness production as compared to proton-proton or
proton-nucleus collisions is one of the QGP signals proposed a
long time ago. In the hydrodynamic model described above only
baryon charge rather than strangeness exchange is included. So, to
see experimental consequences of EoS with different phase
transition order,  we consider an  expanding homogeneous blob of
the compressed  and heated QCD matter  (a fireball) formed in
heavy-ion collisions. The initial state ($\varepsilon_0$ and
$n_B$) for this fireball  is estimated from results of the QGSM
transport calculations in the center-of-mass frame inside a
cylinder of the volume $V_0$ with radius $R=5 \ fm$ and
Lorentz-contracted length $L=2R/\gamma_{c.m.}$. Isoentropic
expansion is treated in an approximate manner assuming $V \sim V_0
t$ until  the freeze-out point defined by $\varepsilon_f = 0.15\
GeV/fm^3 \approx m_N n_0$ (for more detail see~\cite{FNNRT02}).

One should note that till this point the grand canonical ensemble
was used where complete chemical equilibrium is assumed and the
strangeness conservation is controlled on average by the strange
chemical potential $\mu_S$. In the thermodynamical limit,
fluctuations in a number of strange particles are small and
coincide with those for the canonical ensemble. However, it is not
the case for finite systems at relatively small $T$ where the
strangeness canonical ensemble should be applied, taking into
account the associative nature of strange particle creation
 by exact  and local conservation of  strangeness. Using the general
 formalism for the canonical strangeness
conservation proposed in~\cite{HR85,CRS91}, the partition function
of a gas of hadrons with strangeness $s_i =0, \pm 1, \pm 2, \pm3$
and total strangeness $S=0$  can be written as follows
\begin{equation}
Z_S=\frac{1}{2\pi} \int_{-\pi}^{\pi}d\phi \  \exp(\sum_{s=-3}^{3}
{\cal S}_s \ e^{is\phi}) \label{eq26}
\end{equation}
where ${\cal S}_s=V\sum_i Z_i~. $ Here $Z_i$ is the one-particle
partition function for species $i$ and the sum is taken over all
particles and resonances carrying strangeness $s_i$. A number of
strange particles can be found by the appropriate differentiation
of the partition function $Z_S$, Eq.(\ref{eq26}). It is easy to
see that canonical result can be obtain in the Boltzmann
approximation from the grand canonical one by replacing the
strange fugacity in the following way~:
\begin{equation}
 \exp (\mu_s/T) \to  \left(\frac{{\cal S}_1}{\sqrt{{\cal S}_1 {\cal
       S}_{-1}}}\right)^s \ \frac{I_s(x)}{I_0(x)}~,
%\exp (\mu_s/T) \to \frac{{\cal S}_1^s}{({\cal S}_1 {\cal
%    S}_{-1})^{s/2}} \ \frac{I_s(x)}{I_0(x)}~,
\label{eq31}
\end{equation}
where the argument of the Bessel functions $x\equiv 2\sqrt {{\cal
S}_1{\cal S}_{-1}} \sim V$.  This receipt was applied to our
treatment of particle abundance at the freeze-out point.
Generally, the correlation volume in the suppression factor
$I_s(x)/I_0(x)$ of (\ref{eq31}) does not coincide with the system
volume $V$. In our model, the initial Lorentz-contracted volume
$V_0$ is considered as a strangeness correlation volume.
\begin{figure}[hbt]
\begin{center}
\includegraphics[width=10.3cm,clip]{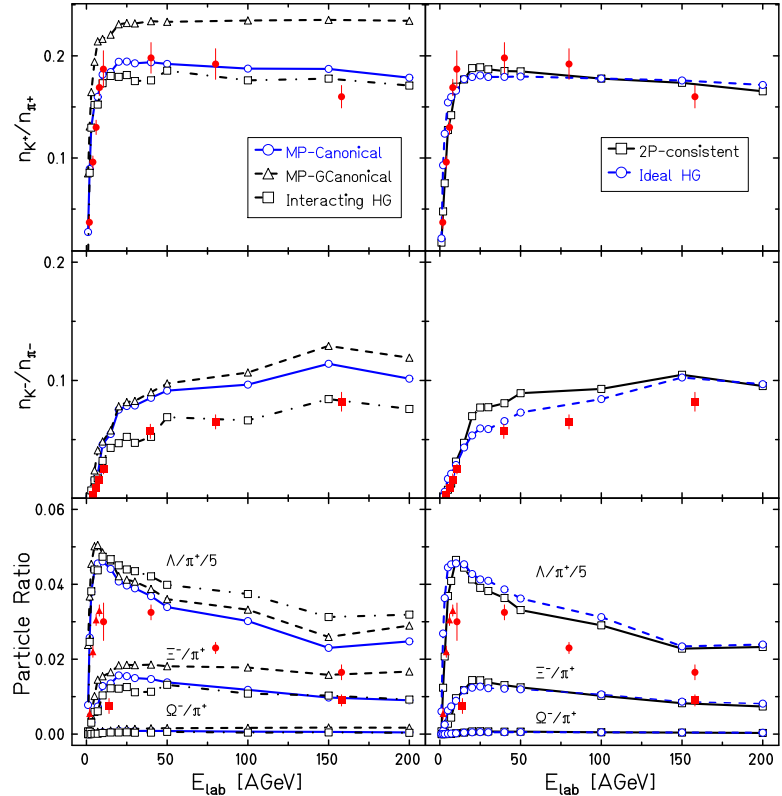}
  \vspace{-4mm}

 \caption[C12]{  Particle ratios for strange hadrons
in full $4\pi$ angle interval for
  central $Au+Au$ collision as a function of bombarding energy. The
  compilation of available experimental points is taken
  from~\cite{Redlich01,Stock02}.
 The calculated excitation functions represent four modeling EoS with
 the canonical suppression factor $I_s(x)/I_0(x)$.
 For the case of the MP model, the
 grand canonical results (dashed lines) are given, as well.}
\label{fig4}
\end{center}
\end{figure}

Inspection of Fig.\ref{fig4} shows that the inclusion of the
canonical strangeness suppression factor  allows one to decrease
noticeably strange particle abundance. However, comparing
canonical and grand canonical results for the MP model, one can
see that they do not coincide at high energies as it would be
expected. This is explained by the beam energy dependence of the
strangeness correlation volume in contrast with usual canonical
description~\cite{HR85,CRS91}. The most striking result followed
from Fig.\ref{fig4} is that all the models considered in the
strangeness canonical ensemble predict practically the same
relative abundance of strange hadrons in the whole energy range
studied. The measured excitation function for $K^+/\pi^+$  are
reproduced  reasonably well excluding maybe the SIS energy. In the
case of $K^-/\pi^-$ the general form of excitation functions also
agrees with experimental one but the relative abundance is
overestimated what mainly originates from neglecting the electric
charge (isospin) conservation. A simple estimate shows that taking
into account the isospin conservation the $K^-/\pi^-$ ratio
decreases by about $18\%$ and $12 \%$ at $E_{lab}=10$ and $150 \
AGeV$, respectively, without any essential influence on the
$K^+/\pi^+$ ratio. The relative yield of hyperons, in particular
$\Lambda/\pi^+$'s, seems to be overshot in the energy range
$E_{lab} \lsim  \ 10 \ AGeV$ what can result from the simplified
dynamical treatment: The Bjorken-like longitudinal expansion can
be applied at the SPS energies, but at the SIS energies the
transverse expansion is not negligible. It is worthy to note that
all calculations have been done with the same shock-like
freeze-out condition  for every EoS without any special tuning.

\section{Conclusions}
It has been shown that the directed flow excitation functions
$F_y$ and $\left< P_x\right>$ for baryons
 are sensitive to the EoS, but this sensitivity
is significantly masked  by nonequilibrium dynamics of nuclear
collisions. Nevertheless, the  results indicate that the widely
used two-phase EoS, based on the  bag model~\cite{HS95,R96} and
giving rise to a first-order phase transition, seems to be
inappropriate. The  neglect of interactions near the deconfinement
temperature results in an unrealistically  strong softest-point
effect within this two-phase EoS. Smooth experimental excitation
function of the directed flow is reasonably reproduced within the
MP model.

The dynamical trajectories for a fireball state in the $T-\mu_B$
plane are quite different for different EoS~\cite{FNNRT02,TCN01}.
However, after using the shock-like freeze-out, the global
strangeness production is turned out to be completely insensitive
to the particular EoS as illustrated by the calculation of
excitation functions for $K^+, K^-$ and hyperons. To get agreement
with experiment the canonical suppression factor should be taken
into account. The only trace of dynamics is the beam-energy
dependence of the strangeness correlation volume. This effect
results in a negative slope of the $K^+/\pi^+$ excitation function
at high energies which is not reproduced  by the equilibrium
statistical model with canonical account of
strangeness~\cite{CRS91,Redlich01}.

Among other signals of the QCD deconfinement phase transition, the
dilepton and hard photon production is the most promising. Being
sensitive to the whole evolution time of a system, these signals
can become experimentally observable to disentangle  the crossover
phase transition, predicted by the MP model, from  the first order
one which is peculiar for the bag-model EoS.

\vspace*{5mm} Useful and numerous discussions with B.~Friman,
Yu.B.~Ivanov, E.G.~Nikonov, W.~N\"{o}ren\-berg and K. Redlich are
acknowledged. This work was supported in part by DFG (project 436
RUS 113/558/0) and RFBR (grant 00-02-04012).

\end{document}